\documentclass[twocolumn]{aastex62}

	\usepackage{graphicx}
	\usepackage{float}
	\usepackage{savesym}
	\savesymbol{tablenum}

	\usepackage{amsmath}

	\usepackage{gensymb}

	\usepackage{siunitx}
	\DeclareSIUnit\groove{(grooves)}

\begin{document}


	\title{Spectra of Ni V and Fe V in the Vacuum Ultraviolet}

	\author{J. W. Ward}
	\affil
		{
			National Institute of Standards and Technology \\
			Gaithersburg, MD 20899-8422, USA
		}
	\affil
		{
			University of Maryland \\
			College Park, MD, 20742, USA
		}

	\author{A. J. J. Raassen}
	\affil
		{
			SRON Netherlands Institute for Space Research
			Utrecht, The Netherlands
		}

	\author{A. Kramida}
	\affil
		{
			National Institute of Standards and Technology \\
			Gaithersburg, MD 20899-8422, USA
		}

	\author{G. Nave}
	\affil
		{
			National Institute of Standards and Technology \\
			Gaithersburg, MD 20899-8422, USA
		}


	\begin{abstract}

			This work presents 97 remeasured Fe V wavelengths
			(\SIrange{1200}{1600}{\angstrom}) and 123 remeasured
			Ni V wavelengths (\SIrange{1200}{1400}{\angstrom}) with
			uncertainties of approximately \SI{2}{\milli\angstrom}.
			An additional 67 remeasured Fe V wavelengths and 72 remeasured
			Ni V wavelengths with uncertainties greater than \SI{2}{\milli\angstrom}
			are also reported. A systematic calibration error is also identified in
			the previous Ni V wavelengths and is corrected in this work.
			Furthermore, a new energy level optimization of Ni V
			is presented that includes level values as well as
			Ritz wavelengths. This work improves upon the available
			data used for observations of quadruply ionized
			nickel (Ni V) in white dwarf stars. This compilation
			is specifically targeted towards observations of the G191-B2B white dwarf
			spectrum that has been used to test for variations in the fine
			structure constant, $\alpha$, in the presence of strong gravitational
			fields \citep{Berengut-Alpha-Dwarf}. The laboratory wavelengths for
			these ions were thought to be the cause of inconsistent conclusions
			regarding the variation limit of $\alpha$ as observed through the white
			dwarf spectrum. These inconsistencies can now be addressed with the improved
			laboratory data presented here.

	\end{abstract}



	\section{Introduction}\label{introduction}

		The development of unification theories that depend upon
		spatial and temporal variations of physical constants has
		and continues to be of interest to the physics community.
		Variations in the fine structure constant, $\alpha$, contribute
		to multiple cosmological models and string theories, as discussed
		by \citet{Martins-Alpha}, such as the Bekenstein-Sandvik-Barrow-Magueijo theory
		\citep{Sandvik-Alpha, Barrow-Alpha, Bekenstein-Alpha}.
		The search for variations in $\alpha$ has previously made
		use of methods involving both measurements based on atomic
		clocks \citep{Berengut-Alpha-Clocks, Blatt-Alpha-Clocks,
		Bauch-Alpha-Clocks} and on the observations of quasar
		spectra \citep{Webb-Alpha-Quasars, Dzuba-Alpha-Quasars}
		with the objective being ever finer limits on the potential variation.

		The motivation behind our work stems from a recent
		publication that investigates the possible dependence
		of $\alpha$ on strong gravitational fields
		\citep{Berengut-Alpha-Dwarf}. The study makes use of
		far-UV spectral observations of Fe V and Ni V in the
		atmosphere of the G191-B2B white dwarf star \citep{Preval-G191-B2B}.
		G191-B2B provides data for an analysis of the fine
		structure constant where the ions producing the observed
		spectrum experience a gravitational potential (relative
		to laboratory conditions) that is five orders of
		magnitude larger than in previous studies based on
		atomic clocks in Earth bound satellites. The analysis conducted
		by \citet{Berengut-Alpha-Dwarf}, however, resulted in
		conflicting estimates for $\Delta \alpha / \alpha$, which is demonstrated
		by Figures 1 and 2 of their paper. The laboratory
		wavelength standards for both Fe V and Ni V dominate the
		uncertainty of the fine structure variation.

		The wavelength values used by \citet{Berengut-Alpha-Dwarf} for Fe V were reported
		by \citet{Ekberg-Fe-V}. The reported wavelengths had estimated
		uncertainties of \SI{4}{\milli\angstrom}. This estimate of the
		wavelength uncertainty is supported by \citet{Berengut-Alpha-Dwarf}.
		Of the wavelengths reported by Ekberg, 96 were used in the investigation
		of fine structure variation covering a wavelength range of approximately
		\SIrange{1200}{1600}{\angstrom}.

		In addition to the report by Ekberg, a rigorous assessment
		and optimization of Fe V data has been conducted by \citet{Kramida-Ritz}.
		\citeauthor{Kramida-Ritz} verified the uncertainty estimate given by
		\citet{Ekberg-Fe-V} and used Ekberg's data in conjunction with data
		from other researchers to derive a set of Fe V Ritz wavelengths
		with uncertainties of \SI{2}{\milli\angstrom} or less.

		The wavelengths for Ni V were reported by Raassen, van Kleef, \& Metsch
		(\citeyear{Raassen-Ni-V-Short}, hereinafter \citeauthor{Raassen-Ni-V-Short})
		and Raassen and van Kleef (\citeyear{Raassen-Ni-V-Long}, hereinafter
		\citeauthor{Raassen-Ni-V-Long}). The reported wavelengths in the
		\SIrange{200}{400}{\angstrom} range had estimated uncertainties of
		\SI{1}{\milli\angstrom}, but the wavelengths in the \SIrange{900}{1400}{\angstrom}
		range were not reported with uncertainties. The uncertainties used in the report
		by \citet{Berengut-Alpha-Dwarf} indicate that a \SI{7}{\milli\angstrom} uncertainty
		seems to be appropriate. This estimated uncertainty is consistent with Raassen's
		report on Ni VI \citep{Raassen-Ni-VI} that gives \SI{6}{\milli\angstrom} as the
		estimated uncertainty in the \SIrange{900}{1300}{\angstrom} range using the same
		calibration method as the one he used for Ni V. Of the wavelengths reported by
		\citeauthor{Raassen-Ni-V-Long}, 32 were used in the investigation of fine structure
		variation covering a wavelength range of approximately \SIrange{1200}{1400}{\angstrom}.


	\section{Experimental Methods}\label{methods}


				\begin{figure*}[htb!]

					\centering

					\epsscale{.75}

					{\plotone{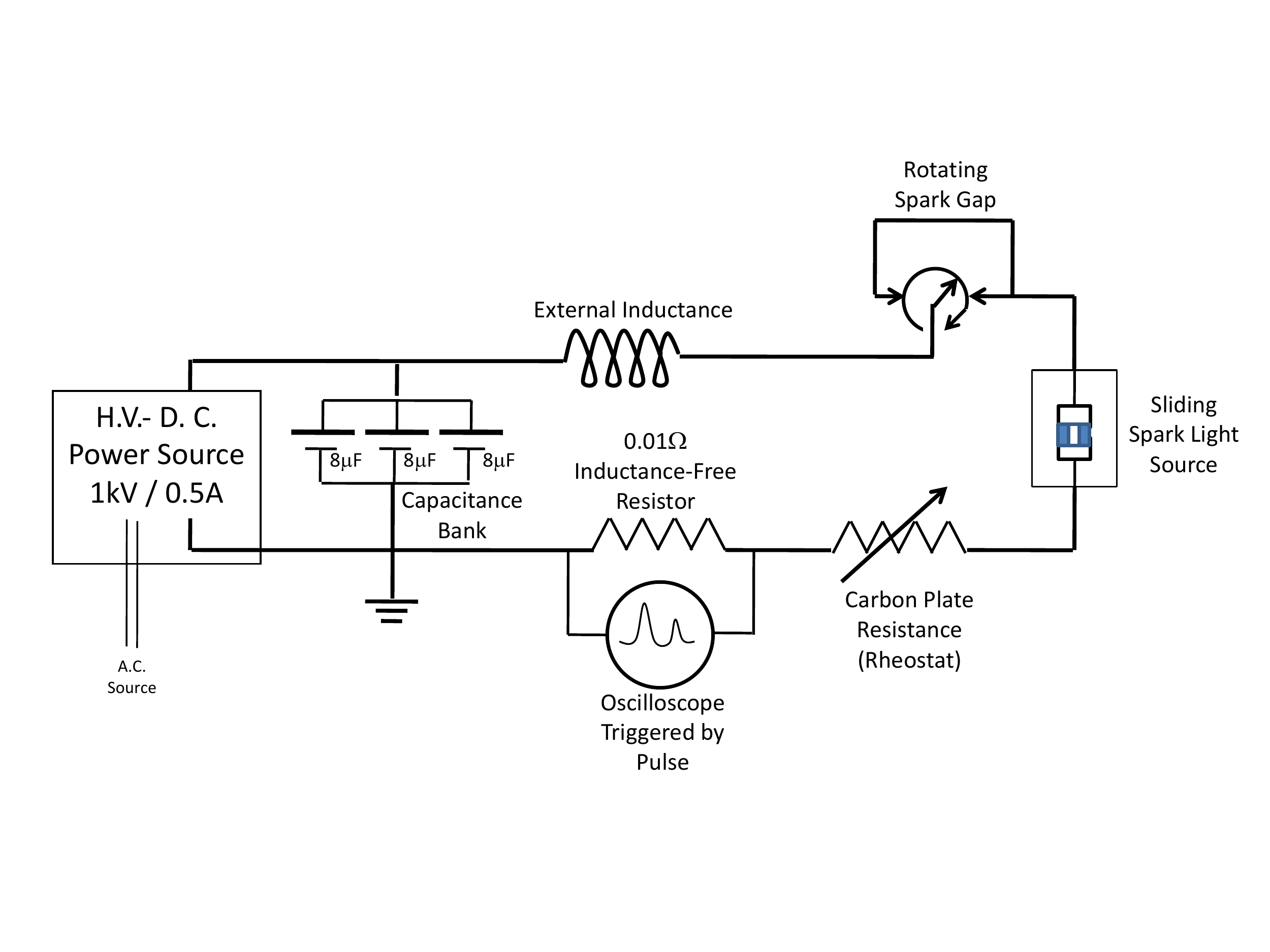}}

					\caption
					{
						A diagram of the circuitry for the sliding spark
						light source.\label{Spark-Circuit}
					}

				\end{figure*}


		The wavelengths in this work were measured with the National
		Institute of Standards and Technology (NIST) \SI{10.7}{\meter}
		Normal Incidence Vacuum Spectrograph (NIVS), which operates
		in the \SIrange{300}{5000}{\angstrom} range. The NIVS is in
		a Rowland Circle configuration that has a focal length of
		\SI{10.7}{\meter} and contains a gold coated, concave grating
		blazed for \SI{1200}{\angstrom} with \SI{1200}{\groove\per\milli\meter}.
		This results in a reciprocal linear dispersion $\approx$ \SI{.78}{\angstrom\per\milli\meter}.
		The image recorded at the plate holder of the NIVS is created by a
		single slit with a width of \SI{21}{\micro\meter}.

		The Ni V and Fe V spectra were obtained with a sliding spark
		source \citep{Vodar-Sliding-Spark, Beverly-Surface-Sparks,
		Reader-Spark-Operation}. A diagram of the circuitry for the
		sliding spark is given in Figure \ref{Spark-Circuit}.
		For this work we have used invar, an iron and nickel alloy, for
		both electrodes in the source. Invar was chosen in order to create
		exposures with both Ni V and Fe V in the same track. This allowed
		Ni V and Fe V to be placed on the same wavelength scale and ensured that
		any systematic errors in the	 calibration are common to both species.
		The exposures analyzed here were taken at a range of peak currents from
		\SI{300}{\ampere} to \SI{2000}{\ampere} with the best spectrum of Ni V and Fe V
		observed at a peak current of \SI{1500}{\ampere}.
		In order to achieve that peak current, the inductor, shown in Figure \ref{Spark-Circuit},
		was removed from the spark circuitry. The carbon plate resistor contained
		thirteen carbon plates, the supply voltage was approximately
		\SIrange{600}{700}{\volt} depending on the given exposure,
		the circuit spark gap was run at a repetition rate of
		\SI{20}{\milli\second}, and the resulting pulse width was
		\SI{50}{\micro\second}. The exposures were run for twenty minutes,
		and the average current was roughly \SI{.5}{\ampere}.

			\begin{deluxetable*}{ccccccc}[t!]
				\tabletypesize{\scriptsize}
				\tablecaption{Table of Spectra\label{Spectra-Table}}
				\tablehead
				{
					\colhead{Plate Number} &
					\colhead{Exposure Date} &
					\colhead{Plate Type$^{a}$} &
					\colhead{Track Number} &
					\colhead{Source$^{b}$} &
					\colhead{Source Conditions} &
					\colhead{$\lambda$ Range ($\AA$)}
				}

					\startdata
						x988 & 07/03/2014 & PIP & 5 & $D_{2}$ & \SI{300}{\milli\ampere}  & \SIrange{1150}{1450}{\angstrom} \\
						x990 & 07/11/2014 & PIP & 1 & Pt/Ne HCL & \SI{20}{\milli\ampere}, \SI{340}{\volt}  & \SIrange{1150}{1450}{\angstrom} \\
						x990 & 07/11/2014 & PIP & 4 & Invar SS & \SI{1000}{\ampere} Peak, \SI{.55}{\ampere} Average, \SI{600}{\volt} & \SIrange{1150}{1450}{\angstrom} \\
						x990 & 07/11/2014 & PIP & 5 & Invar SS & \SI{1500}{\ampere} Peak, \SI{.65}{\ampere} Average, \SI{850}{\volt}  & \SIrange{1150}{1450}{\angstrom} \\
						x997 & 06/04/2015 & KSWR & 1 & Pt/Ne HCL & \SI{20}{\milli\ampere}, \SI{310}{\volt} & \SIrange{1190}{1530}{\angstrom} \\
						x997 & 06/04/2015 & KSWR & 2 & Invar SS & \SI{1500}{\ampere} Peak, \SI{.48}{\ampere} Average, \SI{530}{\volt} & \SIrange{1190}{1530}{\angstrom} \\
						x997 & 06/04/2015 & KSWR & 3 & Fe/Y SS & \SI{1500}{\ampere} Peak, \SI{.45}{\ampere} Average, \SI{750}{\volt} & \SIrange{1190}{1530}{\angstrom} \\
						x997 & 06/04/2015 & KSWR & 4 & Ni/Y SS & \SI{1500}{\ampere} Peak, \SI{.45}{\ampere} Average, \SI{750}{\volt} & \SIrange{1190}{1530}{\angstrom} \\
						x997 & 06/04/2015 & KSWR & 8 & Pt/Ne HCL & \SI{20}{\milli\ampere}, \SI{310}{\volt}  & \SIrange{1190}{1530}{\angstrom} \\
					\enddata

					\tablenotetext{a}
					{
						PIP: Phosphore Image Plate and KSWR: Kodak SWR Photographic Plate
					}
					\tablenotetext{b}{HCL: Hollow Cathode Lamp and SS: Sliding Spark}

			\end{deluxetable*}

\begin{figure*}[htb!]

			\centering

			\plotone{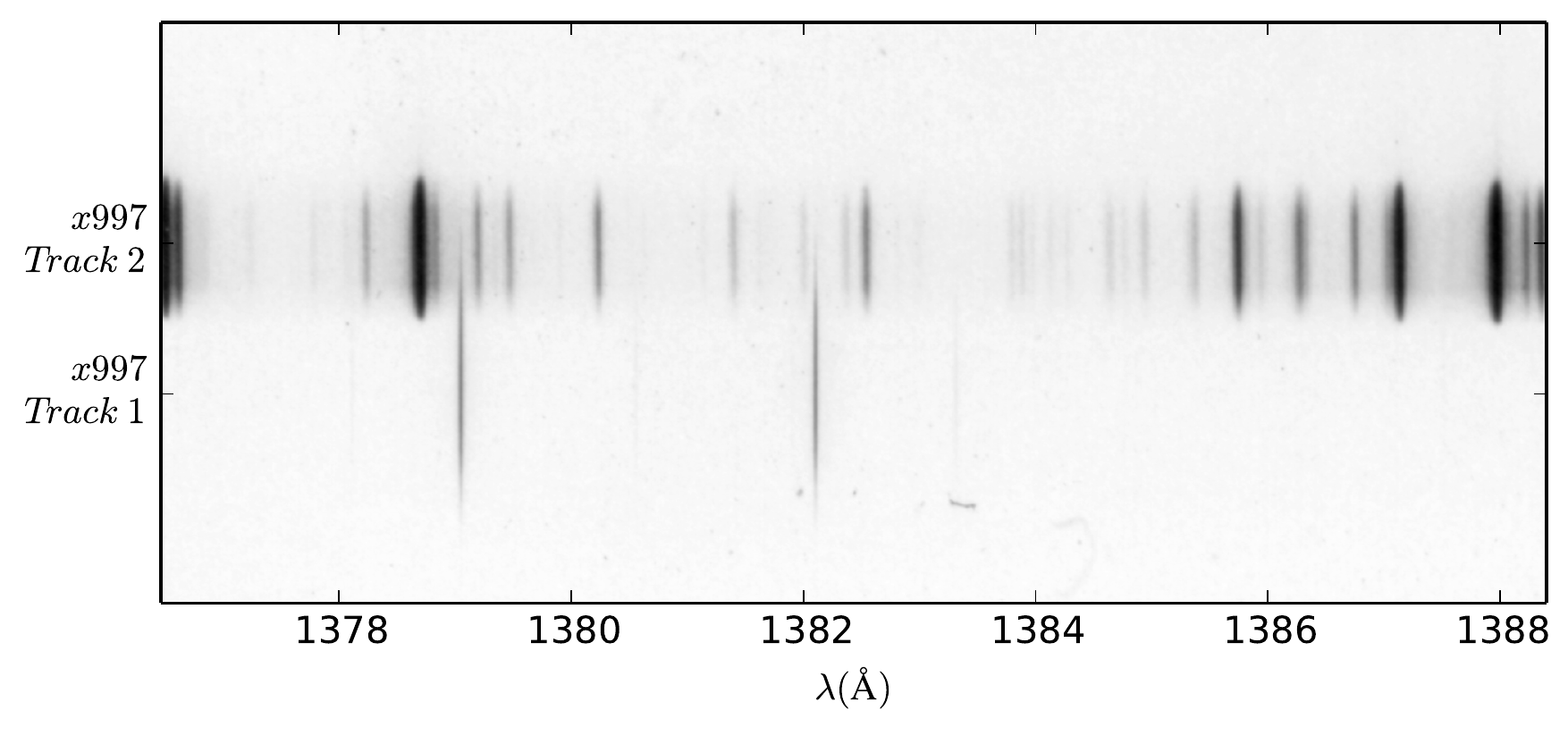}

			\caption
			{
				A sample section from the photographic plate x997, described in Table \ref{Spectra-Table}, that was
				used in our work to measure wavelengths. The top exposure is the spectrum of an invar SS source
				and the bottom exposuire is from a Pt/Ne HCL.\label{Spectrum-Image}
			}

		\end{figure*}

		The spectra were recorded on both Kodak SWR photographic plates\footnote{The identification of
		commercial products in this paper does not imply recommendation or endorsement by
		the National Institute of Standards and Technology, nor does it imply that the items
		identified are necessarily the best available for the purpose.}
		and phosphor image plates. Table \ref{Spectra-Table} presents the
		details for all exposures used in our work, and Figure \ref{Spectrum-Image} shows
		a sample from one of the spectra described in Table \ref{Spectra-Table}.
		The grain size in the photographic plates, roughly \SI{.5}{\micro\meter} \citep{McCrea-SWR-Grain},
		gives the photographic plates a significant advantage over the other
		available VUV imaging techniques in terms of resolution and subsequent linewidth.
		The high density of spectral lines present in the invar spectrum makes the
		additional resolution provided by photographic plates necessary. Attempts
		to develop an accurate set of wavelengths with
		other imaging techniques, such as phosphor image plates, using the XGREMLIN
		software \citep{XGREMLIN}, were hindered by a significant number of blended
		lines.


		The wavelength scale for the invar spectrum recorded on photographic plates
		was calibrated with a Pt II spectrum produced with a platinum/neon hollow-cathode lamp (HCL)
		run with a current of \SI{20}{\milli\ampere}. The Pt II spectrum was partially embedded
		in the invar spectrum without moving the photographic plates
		between the platinum and invar exposures (shown in Figure \ref{Spectrum-Image}).
		This was done in order to eliminate the effects of moving the plates between the calibration
		spectrum and experimental spectrum. Attempts to apply a calibration
		derived from a separate track than the invar spectrum,
		that required vertically translating the plates, yielded a linear slope of spectrum
		along the plate. The sloping effects we observed are likely due
		to a tilting of the plates during the process of moving them vertically
		between separate exposures. When the plates were not moved between
		separate exposures the calibrated wavelengths of contaminant lines in the invar
		spectrum and the Fe V wavelengths that had available Ritz wavelengths were in much
		better agreement with their reported values.

		The positions of the spectral lines present on the photographic
		plates were measured using the NIST rotating mirror comparator \citep{Tomkins-Comparator}.
		The measurement uncertainty associated with the use of the NIST
		comparator was evaluated by taking multiple measurements of the same set of 83
		well measured lines present in the invar spectrum and taking the
		standard deviation of the calibrated wavelengths that resulted
		from the line position measurements. The standard uncertainty
		introduced by the comparator measurement was determined to be
		\SI{2}{\milli\angstrom} for lines without serious perturbations such as an
		asymmetry or blend. Lines with perturbations such as asymmetry or
		blending were given an increased measurement uncertainty, ranging from
		an additional \SI{1}{\milli\angstrom} to \SI{10}{\milli\angstrom}, corresponding to
		the impact of the perturbation.

		The radiometric calibration of the invar spectrum recorded on phosphor image plates
		was done with a deuterium standard lamp that was calibrated at the Physikalish-Technische
		Bundesanstalt (PTB). The $D_{2}$ spectrum, as well as the invar spectrum,
		was recorded on phosphor image plates for the radiometric calibration. We chose phosphor
		image plates for the radiometric calibration because they scale linearly with intensity
		\citep{Nave-Image-Plates}, unlike the photographic plates which have a non-linear response
		in intensity.


	\section{Analysis}\label{analysis}

		\subsection{Calibration}\label{calibration}

			\subsubsection{Wavelength Calibration}\label{wave-calibration}

				The wavelength calibration for the Ni V and Fe V spectra was
				carried out with Pt II reference wavelengths \citep{Pt-II-Atlas}.
				Of the 93 platinum reference values used in the calibration of
				the invar spectrum, 59 of the wavelengths have uncertainties of
				less than \SI{2}{\milli\angstrom} with the remaining values having
				uncertainties of \SI{2}{\milli\angstrom}.


				\begin{figure}

					\centering

					\plotone{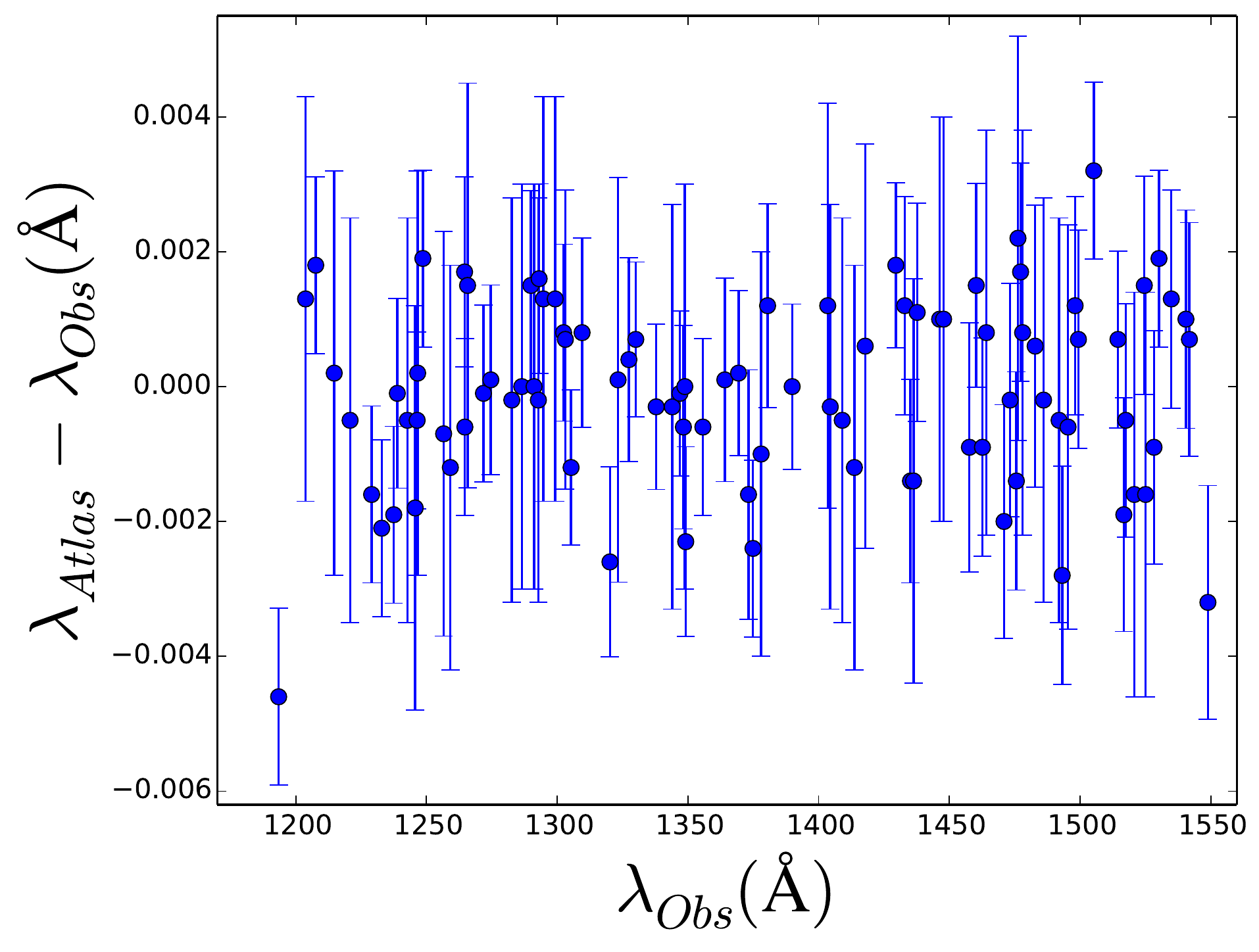}

					\caption
					{
						Residuals after fitting our observed Pt II wavelengths ($\lambda_{Obs}$)
						with a sixth order polynomial to their Pt II standard reference wavelengths
						($\lambda_{Atlas}$) \citep{Pt-II-Atlas}.\label{Pt-II-Graph}
					}

				\end{figure}


				The calibration function was created by identifying the positions on
				photographic plates of Pt II lines in the Pt/Ne spectrum that had
				wavelengths from \citet{Pt-II-Atlas}. The line positions and reference
				wavelengths were then used to derive a dispersion function that was a
				sixth order polynomial. Once the dispersion function was determined,
				the line positions of Fe V, Ni V, and contaminant lines were measured
				and the polynomial dispersion function was applied. The contaminant
				lines, such as Y IV \citep{Epstein-Y-IV} and Si IV \citep{Griesmann-Si-IV},
				were then used to identify and correct illumination shifts between
				the calibration source (Pt/Ne HCL) and the experimental source (invar SS).

		 		The standard uncertainty introduced due to the calibration, estimated
				by the standard deviation of the calibration residuals shown in Figure
				\ref{Pt-II-Graph}, is \SI{1.3}{\milli\angstrom}.

			\subsubsection{Intensity Calibration}\label{intensity-calibration}

				With the $D_{2}$ spectrum discussed in section \ref{methods} we established
				an accurate intensity scale and report relative intensities for the
				observed Ni V lines. The approach to the radiometric calibration follows the same
				procedure discussed in section IV, subsection A of \citet{Nave-Image-Plates}. Since
				the $D_{2}$ spectrum below \SI{1660}{\angstrom} consists of emission lines,
				the peak intensity of the lines depends on the resolution of the spectrograph.
				As the resolution of the spectrograph used at PTB to calibrate the $D_{2}$ lamp
				was much lower than ours, we degraded our measured spectrum by convolving it with two
				boxcar functions of width \SI{9.2}{\angstrom} and \SI{4.6}{\angstrom} to match the
				resolution of the spectrograph used by PTB. We then interpolated the calibration provided by PTB
				to the same wavelength scale as our degraded spectra and took the ratio of the two spectra to
				create an instrument response function.

				The instrument response function derived from this process was then applied to the invar
				spectrum by taking the ratio of the instrument response function and the invar spectrum signal.
				Each spectral line in the calibrated spectrum was fitted with a Voigt profile using the
				Xgremlin program \citep{XGREMLIN}. The peak value of the Voigt profile was taken as the line
				intensity of the spectral line.

				The estimated uncertainty of the radiometric calibration is \SI{12}{\percent} and
				was derived in a way that is similar to the uncertainty budget described in section IV,
				subsection B of \citet{Nave-Image-Plates}. This uncertainty is a summation in
				quadrature of the \SI{10}{\percent} uncertainty due to variations in the
				alignment of the source and the \SI{7}{\percent} uncertainty that comes from the supplied calibration
				of the $D_{2}$ lamp from PTB. Since the line intensities are highly dependent on the source
				conditions and illumination, they are provided here as only a guide to the spectrum. Caution
				and great care should be used if the intensities are used for other purposes such as for
				calculating transition probabilities.

		\subsection{Wavelength Analysis}\label{Ni-V-Wave}


				\begin{figure*}[htb!]

					\centering

					\plottwo{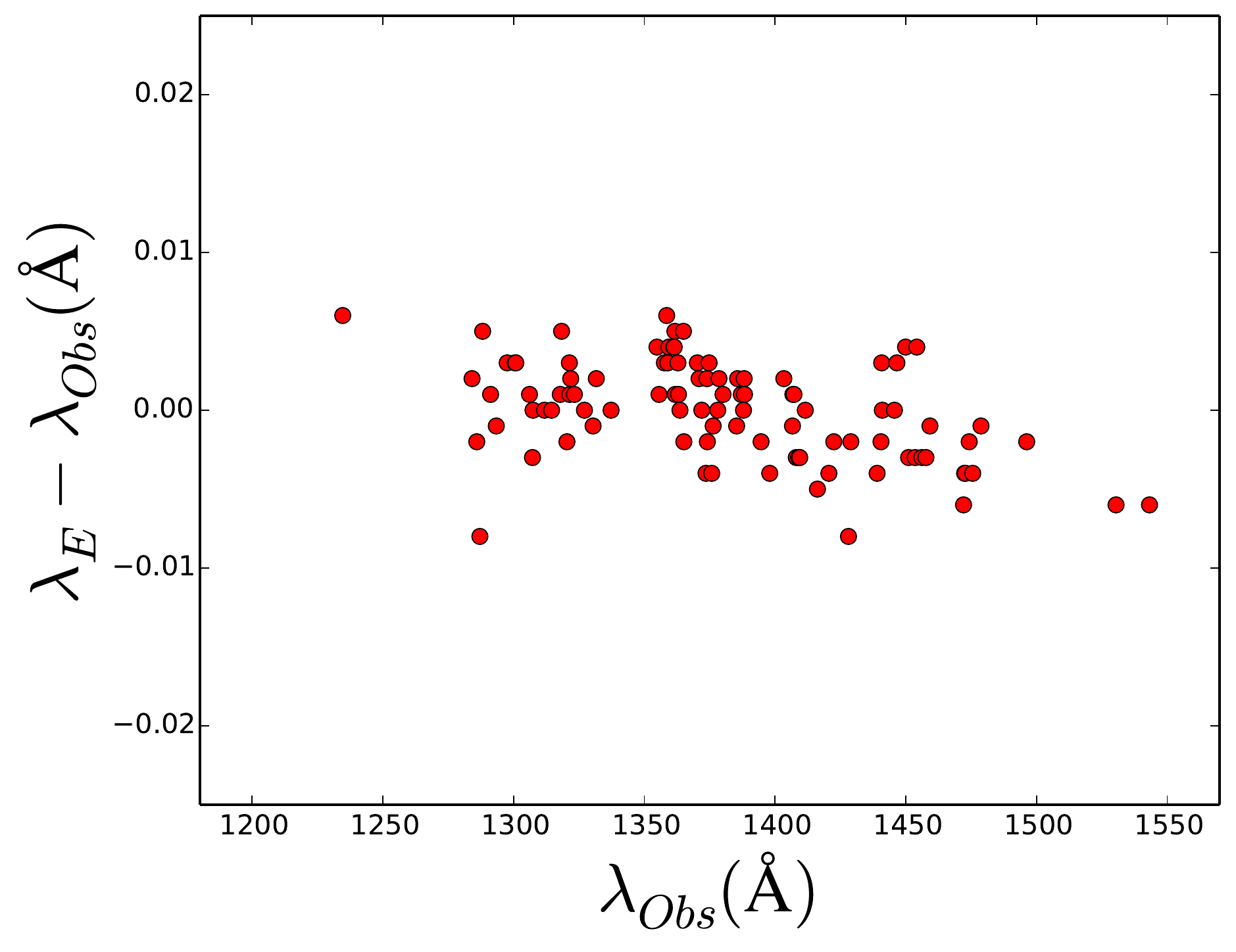}{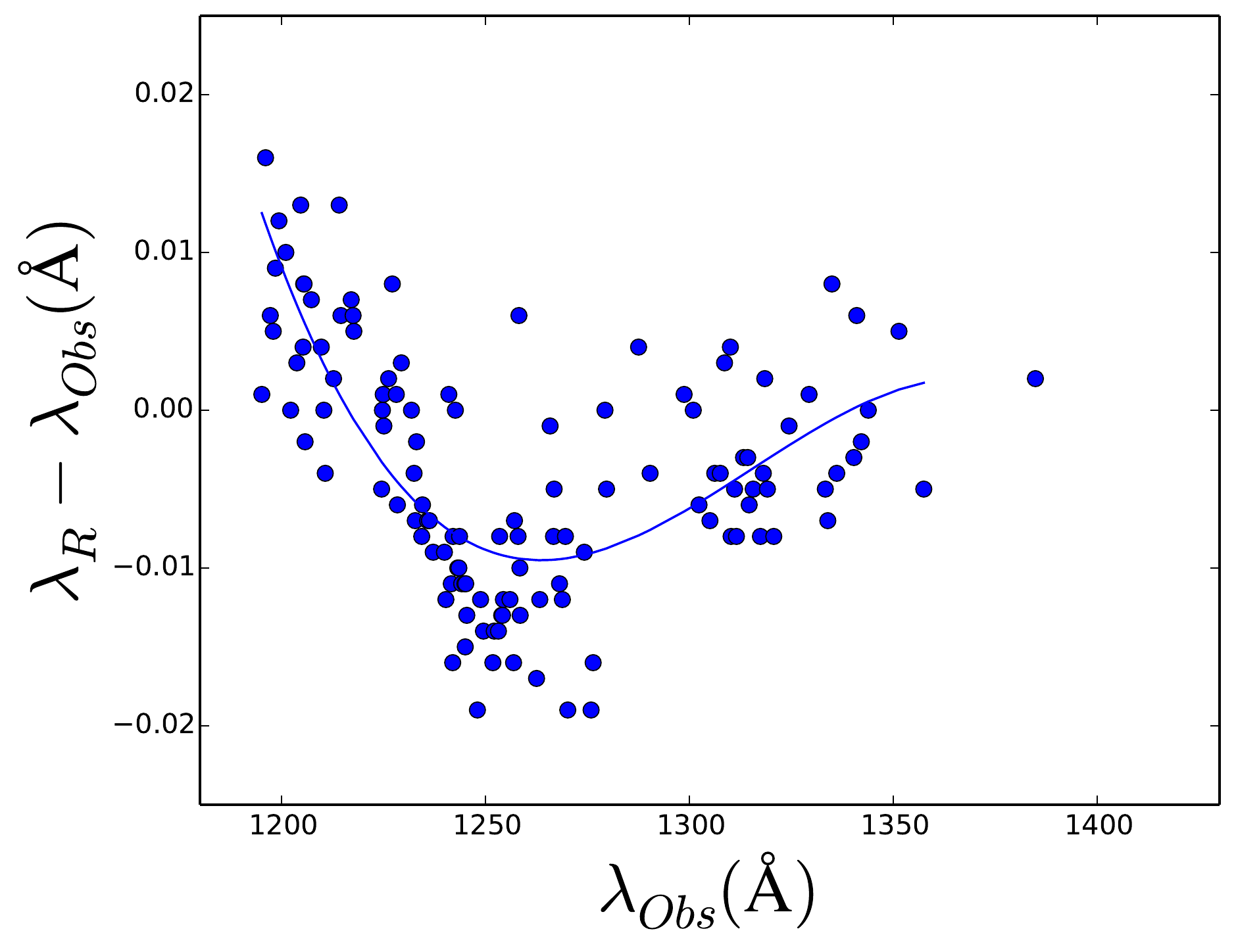}

					\caption
					{
						A comparison of the newly measured wavelengths ($\lambda_{Obs}$) to their previous
						values as reported by either \citet{Ekberg-Fe-V} ($\lambda_{E}$) (Fe V) (Left) or \citeauthor{Raassen-Ni-V-Long}
						($\lambda_{R}$) (Ni V) (Right). The uncertainty of each point is \SI{5}{\milli\angstrom} (Left) and
						\SI{7}{\milli\angstrom} (Right). The Ni V points (Right) are fitted by a third order polynomial shown
						by the solid line.\label{Good-Invar-Graph}
					}

				\end{figure*}


			Figure \ref{Good-Invar-Graph} shows the comparison of the newly
			measured wavelength values to their previously reported values.
			Included are 97 of the 164 observed Fe V wavelengths and 123
			of the 195 observed Ni V wavelengths. All included values
			are from unblended and symmetric lines. We excluded lines that were
			obscured by the calibration spectrum being partially embedded in the invar spectrum
			as a sufficiently accurate measurement of the line position was
			not possible. The excluded values are reported in seperate tables
			(\ref{Ni-V-Table} and \ref{Fe-V-Table}) with an increase in their reported uncertainties
			reflecting their perturbed measurements. The standard deviation
			of the difference in wavelengths shown in Figure \ref{Good-Invar-Graph} for Fe V
			is \SI{3}{\milli\angstrom} and the standard deviation of the difference in
			wavelengths shown in Figure \ref{Good-Invar-Graph} for Ni V is \SI{8}{\milli\angstrom}.

			For the observations of both Ni V and Fe V the reported standard
			uncertainties, \SI{2.4}{\milli\angstrom}, are the sum in quadrature of the calibration
			uncertainty discussed in \ref{wave-calibration} and the
			line position measurement uncertainty discussed in section \ref{methods}.

			An analysis of Figure \ref{Good-Invar-Graph} demonstrates
			the principle improvement found in our work. The Ni V
			comparison shown in the figure highlights a systematic
			difference between the newly measured wavelengths and
			the previous values from \citeauthor{Raassen-Ni-V-Long}. This suggests a systematic
			error in the calibration method used by \citeauthor{Raassen-Ni-V-Long}. This type of
			systematic error has been found previously in reports similar to
			\citeauthor{Raassen-Ni-V-Long}. For example, in the work on the Co III spectrum
			by \citet{Nave-Co-III} a similar trend was observed for wavelengths reported by
			\citet{Raassen-Co-III}.

			The impact of correcting this systematic calibration error,
			concentrated in the \SIrange{1200}{1300}{\angstrom} range,
			should be clear, given that the maximum error introduced by the
			faulty calibration is approximately \SI{10}{\milli\angstrom},
			which would contribute significantly to any application requiring
			Ni V wavelengths. The maximum discrepancy in Figure 2 of
			\citet{Berengut-Alpha-Dwarf} is roughly \SI{12}{\milli\angstrom},
			suggesting that the majority of the discrepancy they observed can by
			explained by the laboratory wavelengths.


		\newpage

			\floattable

			\begin{deluxetable*}{cccccccccccccccc}
				\tabletypesize{\scriptsize}
				\rotate
				\tablecaption{Compilation of Ni V Data\label{Ni-V-Table}}
				\tablehead
				{
					\colhead{$\lambda_{obs}$\,$^{a}$} &
					\colhead{$u_{obs}$\,$^{b}$} &
					\colhead{$\lambda_{Ritz}$\,$^{c}$} &
					\colhead{$u_{Ritz}$\,$^{d}$} &
					\colhead{I\,$^{e}$} &
					\colhead{$\log(gf_{R})$\,$^{f}$} &
					\colhead{$\log(gf_{R})$\,$^{g}$} &
					\colhead{$E_{i}$\,$^{h}$} &
					\colhead{$E_{k}$\,$^{i}$} &
					\multicolumn{3}{c}{Lower Level} &
					\multicolumn{3}{c}{Upper Level} &
					\colhead{Notes\,$^{j}$} \\
					\colhead{{(}\si{\angstrom}{)}} &
					\colhead{{(}\si{\angstrom}{)}} &
					\colhead{{(}\si{\angstrom}{)}} &
					\colhead{{(}\si{\angstrom}{)}} &
					\colhead{Acc.} &
					\colhead{} &
					\colhead{} &
					\colhead{{(}\si{\cm}$^{-1}${)}} &
					\colhead{{(}\si{\cm}$^{-1}${)}} &
					\colhead{Configuration} &
					\colhead{Term} &
					\colhead{J} &
					\colhead{Configuration} &
					\colhead{Term\,$^{k}$} &
					\colhead{J}
				}

\startdata
199.1540 & 0.0019 & 199.1540  & 0.0019 & 41  &        &    &      0.00 & 502124.0  & 3d$^6$              & $^5$D & 4 & 3d$^5$($^6$S)5f     & $^5$F\degree & 5 & R2      \\
199.5040 & 0.0019 & 199.5040  & 0.0019 & 37  &        &    &    889.61 & 502132.7  & 3d$^6$              & $^5$D & 3 & 3d$^5$($^6$S)5f     & $^5$F\degree & 4 & R2      \\
         &        & ...       &        &     &        &    &           &           &                     &       &   &                     &              &   &         \\
1003.233 & 0.022  & 1003.2499 & 0.0031 & 27  & -1.057 & E  & 217049.7  & 316725.76 & 3d$^5$($^4$G)4s     & $^3$G & 5 & 3d$^5$($^2$H)4p     & $^3$G\degree & 5 & R4      \\
1008.269 & 0.022  & 1008.2699 & 0.0056 & 14  & -1.317 & E  & 217100.94 & 316280.73 & 3d$^5$($^4$G)4s     & $^3$G & 3 & 3d$^5$($^2$F$_1$)4p & $^3$F\degree & 3 & d,R4    \\
         &        & ...       &        &     &        &    &           &           &                     &       &   &                     &              &   &         \\
1187.168 & 0.025  & 1187.2012 & 0.0029 &  2  & -1.496 & E  & 235421.42 & 319653.14 & 3d$^5$($^2$F$_1$)4s & $^3$F & 4 & 3d$^5$($^4$F)4p     & $^3$G\degree & 5 & d,R3    \\
1187.770 & 0.025  & 1187.7953 & 0.0048 & 35  & -1.249 & E  & 208164.06 & 292353.65 & 3d$^5$($^4$G)4s     & $^5$G & 4 & 3d$^5$($^4$G)4p     & $^3$H\degree & 5 & p,R3    \\
         &        & ...       &        &     &        &    &           &           &                     &       &   &                     &              &   &         \\
1201.752 & 0.022  & 1201.7557 & 0.0056 & 46  & -1.043 & E  & 263701.48 & 346913.07 & 3d$^5$($^2$D$_2$)4s & $^3$D & 1 & 3d$^5$($^2$D$_2$)4p & $^3$P\degree & 2 & d,bl,R1 \\
1201.849 & 0.022  & 1201.8470 & 0.0084 & 29  & -0.784 & E  & 247165.79 & 330371.06 & 3d$^5$($^2$F$_2$)4s & $^3$F & 2 & 3d$^5$($^2$S)4p     & $^3$P\degree & 1 & p,R1    \\
         &        & ...       &        &     &        &    &           &           &                     &       &   &                     &              &   &         \\
1228.167 & 0.002  & 1228.1685 & 0.0016 & 100 & -0.810 & E  & 242504.64 & 323926.69 & 3d$^5$($^2$G$_2$)4s & $^3$G & 4 & 3d$^5$($^2$H)4p     & $^3$H\degree & 4 & I,W     \\
1228.432 & 0.002  & 1228.4291 & 0.0015 &  20 & -0.129 & E  & 242504.64 & 323909.42 & 3d$^5$($^2$G$_2$)4s & $^3$G & 4 & 3d$^5$($^2$G$_2$)4p & $^3$H\degree & 5 & I,W \\
\enddata

			\tablecomments{Table \ref{Ni-V-Table} is published in its entirety in the electronic edition of the {\it Astrophysical Journal}.}

				\tablenotetext{a}{Experimentally measured wavelengths.}
				\tablenotetext{b}{One standard uncertainty of the wavelength value in the previous column.}
				\tablenotetext{c}{Ritz wavelengths derived from the optimized energy levels calculated by LOPT \citep{Kramida-LOPT}.}

				\tablenotetext{d}
				{
					Estimated uncertainty of the ritz wavelengths reported in the previous
					column. \\ The uncertainty of the ritz wavelength is determined as part
					of the level optimization routine in LOPT \citep{Kramida-LOPT}.
				}

				\tablenotetext{e}{Relative Intensity.}
				\tablenotetext{f}{$\log(gf)$ values calculated by A. J. J. Raassen and P. H. M. Uylings.}

				\tablenotetext{g}
				{
					Estimated uncertainty of the $\log(gf)$ values reported
					in the previous column: \\ B$+$ -- Uncertainty $\leq \SI{7}{\percent}$,
					C$+$ -- Uncertainty $\leq \SI{18}{\percent}$, E -- Uncertainty $> \SI{50}{\percent}$.
				}

				\tablenotetext{h}{Energy of the lower level from the level optimization. See Table \ref{Level-Opt-Table} for energy level uncertainty}
				\tablenotetext{i}{Energy of the upper level from the level optimization. See Table \ref{Level-Opt-Table} for energy level uncertainty}

				\tablenotetext{j}
				{
					Additional line information: d -- diffuse, bl -- blended,
					l -- shaded longer, p -- perturbed by close line,
					w -- wide, s -- shaded shorter, H -- very hazy,
					G -- rough estimate for line position, I -- line
					intensity is unreliable, * -- line is multiply classified,
					z -- transition weighted out of level optimization,
					R1, R2, R3, R4 -- Line intensity and experimental wavelength taken from
					\citeauthor{Raassen-Ni-V-Short} or \citeauthor{Raassen-Ni-V-Long} with
					the wavelength shifted according to section \ref{Ni-V-Wave},
					W -- line information based on measurements made in this work.
				}

				\tablenotetext{k}{\degree -- Odd Parity.}

		\end{deluxetable*}

		\newpage

	\section{Results}\label{results}

			We have combined our work with the corrected wavelengths from \citeauthor{Raassen-Ni-V-Short}
			and \citeauthor{Raassen-Ni-V-Long}, described in section \ref{Ni-V-Wave-Results}, and used
			them to derive optimized energy levels and Ritz wavelengths.
			Table \ref{Ni-V-Table} provides the full results of our compilation for Ni V.
			Columns 1 and 2 give observed wavelengths and their standard uncertainties
			as described in sections \ref{Ni-V-Wave} and \ref{Ni-V-Wave-Results}. Columns
			3 and 4 give Ritz wavelengths, derived from the optimized energy levels, and
			their standard uncertainties as described in section \ref{LOPT}. Column 5 gives
			the relative intensity of the line as described in sections \ref{intensity-calibration}
			and \ref{Intensity-Results}. Columns 6 and 7 give the $\log(gf)$ values of
			each transition as well as the estimated standard uncertainty of each $\log(gf)$
			value as described in section \ref{log(gf)}. Columns 8 and 9 give the lower and
			upper optimized energy levels for each transition as described in section \ref{LOPT}.
			Columns 10 through 15 give the lower and upper configuration, term, and J value of each
			transition. Column 16 provides additional notes for each transition with each note
			character being described in the footer of Table \ref{Ni-V-Table}.

		\subsection{Wavelengths}

			\subsubsection{Fe V}

\begin{figure}[H]

					\centering

					\plotone{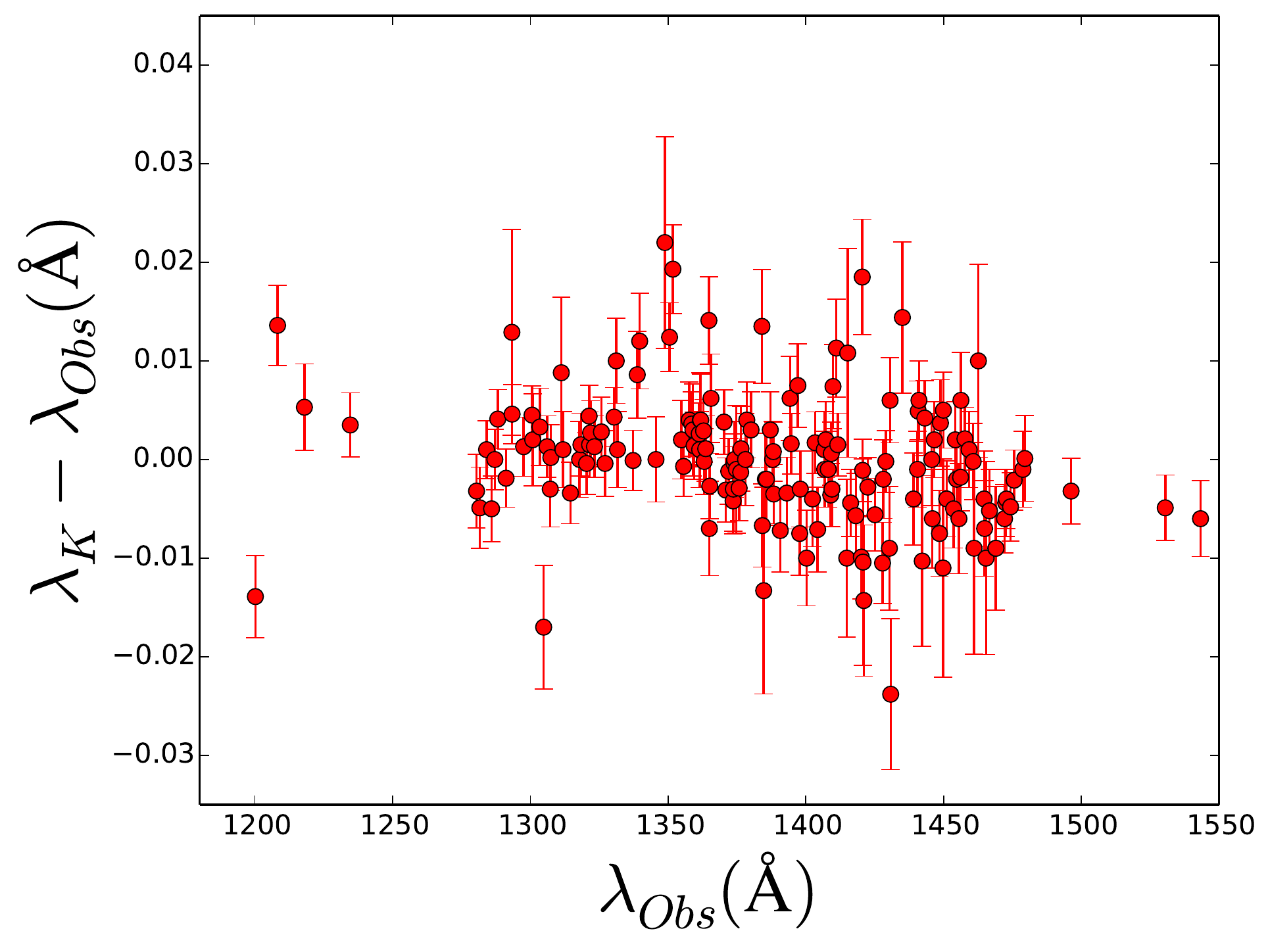}

					\caption
					{
						A comparison of the new Fe V wavelengths ($\lambda_{Obs}$) to the Ritz wavelengths ($\lambda_{K}$)
						from \citet{Kramida-Ritz}.\label{Ritz-Comparison-Graph}
						}

						\end{figure}

				In addition to comparisons with the values from \citet{Ekberg-Fe-V}, we have
				also compared our results to the Ritz values from \citet{Kramida-Ritz}.
				In almost all cases the two sets of wavelengths agree with
				each other to within one standard uncertainty. Overall
				the two reports support each other, which can be clearly seen
				in figure \ref{Ritz-Comparison-Graph}, which shows a standard deviation
				of \SI{7}{\milli\angstrom} in the difference between the two sets. Figure
				\ref{Ritz-Comparison-Graph} does show a small sloping trend in the difference
				between the two sets of wavelengths towards longer wavelengths. This indicates
				that there is still a small systematic error in one of the sets of wavelengths,
				but the sloping trend in figure \ref{Ritz-Comparison-Graph} shows that the
				remaining systematic error is small relative to the wavelength uncertainties.

				Figures \ref{Good-Invar-Graph} and \ref{Ritz-Comparison-Graph} show
				that no substantial improvements have been made for Fe V as a result
				of the new measurements reported in this work. Our
				measurements do, however, validate the wavelengths reported by \citet{Ekberg-Fe-V}
				and \citet{Kramida-Ritz}. Ultimately, the assessment of Fe V by
				\citet{Kramida-Ritz} stands as the recommended source of reference
				data for Fe V.


			\subsubsection{Ni V}\label{Ni-V-Wave-Results}

				The reports by \citeauthor{Raassen-Ni-V-Long} and
				\citeauthor{Raassen-Ni-V-Short} span a wavelength range of
				\SIrange{200}{1400}{\angstrom} and include approximately 1500 spectral lines.
				Roughly 300 of the lines that fall between \SIrange{1200}{1400}{\angstrom}
				were not remeasured in our work. The wavelengths for these lines can
				be corrected by shifting them to the same wavelength scale as the newly
				remeasured wavelengths. This was done by fitting the points in Figure
				\ref{Good-Invar-Graph} with a third order polynomial. The standard deviation
				of the residuals of the third order polynomial fit shown for Ni V in
				figure \ref{Good-Invar-Graph} is \SI{5}{\milli\angstrom}. This was
				then applied to the wavelengths reported by \citeauthor{Raassen-Ni-V-Long}
				in the \SIrange{1200}{1300}{\angstrom} region to give wavelengths on our new scale.
				The wavelengths in Table \ref{Ni-V-Table} that have been corrected in this way
				are reported as the observed wavelengths with a mark (R1) in the note column.

				In the wavelength region that did not overlap with the remeasured wavelengths,
				the accuracy of the wavelength scale can be examined using Ritz wavelengths.
				Accurate relative values of the $3d^{5}4p$ levels were derived from
				$3d^{5}4s$-$3d^{5}4p$ transitions in the \SIrange{1200}{1400}{\angstrom} range using
				the level optimization program LOPT described in more detail in section \ref{LOPT}.
				The relative values and uncertainties of the $4p$ levels are determined solely by
				lines in the \SIrange{1200}{1400}{\angstrom} region. The absolute values are set
				by fixing the  value of one level in the optimization. The $3d^{5}4p$ levels combine
				with each level in the $3d^{6}$ configuration to give transitions in the \SIrange{300}{400}{\angstrom}
				region. The relative Ritz wavelengths and uncertainties of transitions down to a single
				$3d^{6}$ level are determined by lines in the longer wavelength region and can be compared
				to the measured values from \citeauthor{Raassen-Ni-V-Short} to evaluate the accuracy of
				their wavelength scale by looking for systematic deviations. For example, the
				$3d^{6}$ $^{3}P_{2}$ level at \SI{262152}{\per\cm} combines with $3d^{5}4p$ levels to
				give 25 lines from \SIrange{329.25}{382.37}{\angstrom}. A systematic deviation from a
				constant value in the difference between the measured and Ritz wavelengths for these
        lines would suggest a problem in the relative wavelengths in \citeauthor{Raassen-Ni-V-Short}.
        This technique does not validate the absolute wavelength calibration as the absolute values
				of the $4p$ levels must be determined by at least one $3d$-$4p$ transition, but can determine
				if a wavelength calibration error similar to that shown in figure \ref{Good-Invar-Graph} exists
				in the shorter wavelength region.

\begin{figure}[htb!]

					\centering

					\plotone{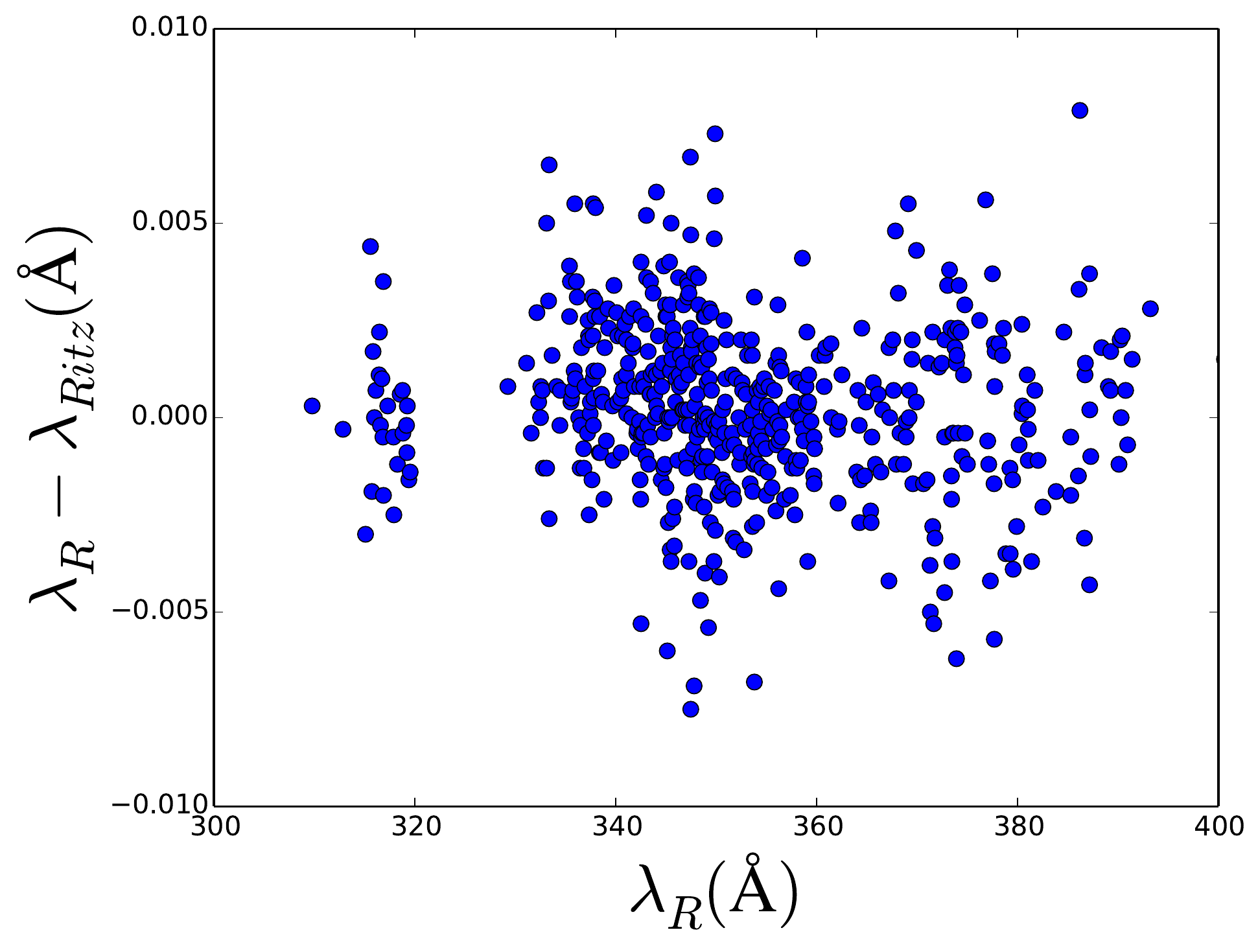}

					\caption
					{
						A comparison of the wavelengths reported by \citeauthor{Raassen-Ni-V-Short}
						($\lambda_{R}$) to the Ritz wavelengths described above
						($\lambda_{Ritz}$).\label{Ritz-Rass-Short-Range}
					}

				\end{figure}

				In our case, it was necessary to fix the values of two $3d^{5}4s$ energy levels in the level
				optimization in order to provide values for a sufficient number of $4p$ levels to determine
				Ritz wavelengths across the whole \SIrange{300}{400}{\angstrom} wavelength range.
				The $3d^{5}(^{4}D)4s$ $^{5}D_{2}$ level was set at \SI{216590.519}{\per\cm} and
				the $3d^{5}(^{2}I)4s$ $^1I_{6}$ level at \SI{233840.023}{\per\cm} using an initial optimization
				of all lines in the \SIrange{200}{1400}{\angstrom} wavelength range. Values for 21 levels in
				the $3d^{6}$ configuration were then fixed using single $3d$-$4p$ transitions and Ritz wavelengths
				for $3d$-$4p$ transitions calculated using the fixed $3d^{6}$ levels and optimized $3d^{5}4p$ levels.
				The results of this comparison, shown in Figure \ref{Ritz-Rass-Short-Range}, indicate no calibration
				error in the \SIrange{300}{400}{\angstrom} range as there is no systematic behavior, and the
				scatter in the difference between the wavelengths is within the estimated measurement uncertainty.
				From this assessment we have reported the original wavelengths in Table \ref{Ni-V-Table} in
				the \SIrange{300}{400}{\angstrom} range given by \citeauthor{Raassen-Ni-V-Short} with a mark
				(R2) in the note column.


\begin{figure}[htb!]

					\centering

					\plotone{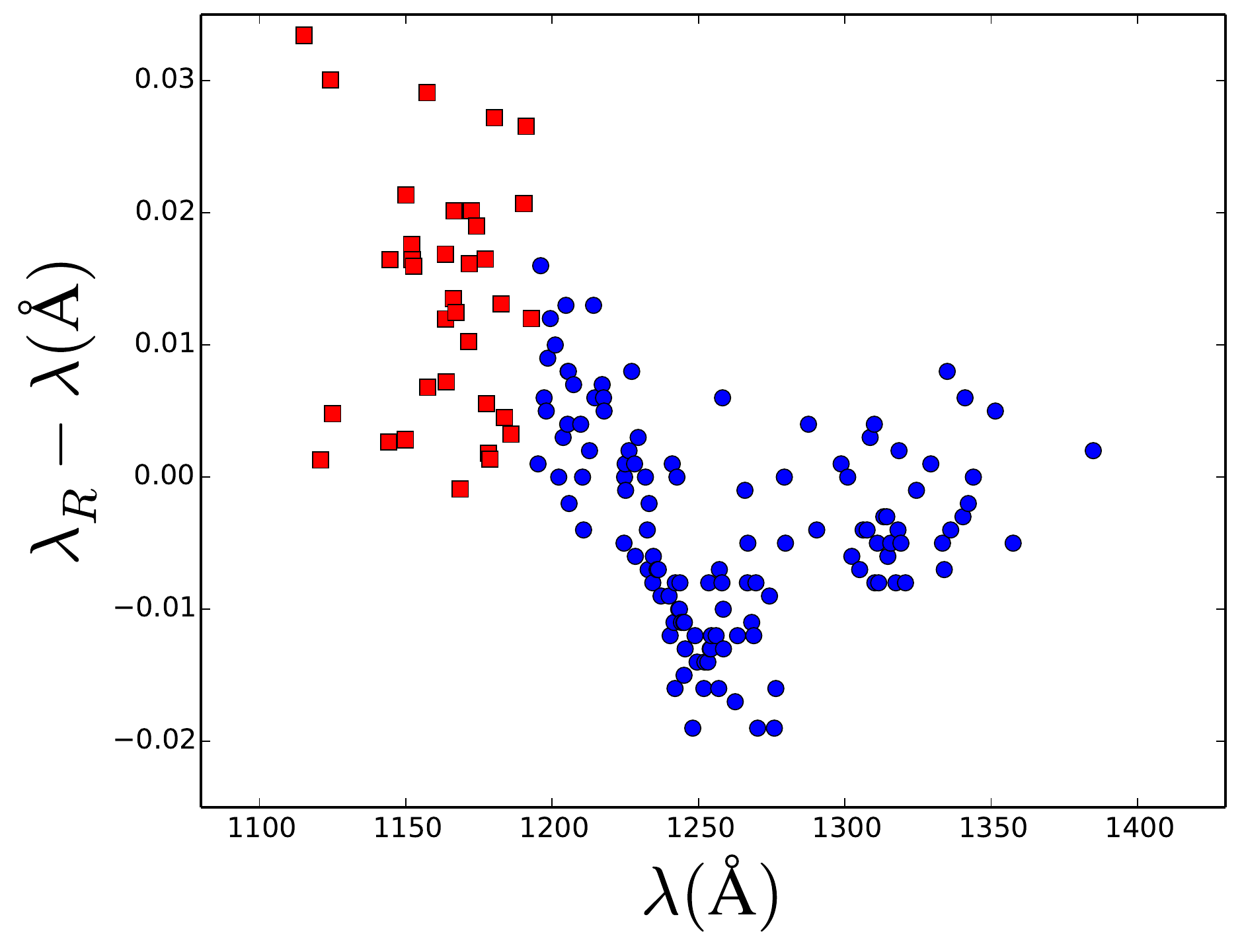}

					\caption
					{
						A comparison of the wavelengths reported by \citeauthor{Raassen-Ni-V-Long}
						($\lambda_{R}$) to our newly measured wavelengths ($\lambda$) (blue circles) and to the
						Ritz wavelengths described in section \ref{Ni-V-Wave-Results} ($\lambda$) (red squares).
						The two comparisons, when joined together, show that the calibration error described in
						section \ref{Ni-V-Wave} extends into the shorter wavelength region shown with the red
						squares.\label{Ritz-Rass-Mid-Range}
					}

				\end{figure}

				In the \SIrange{900}{1200}{\angstrom} range, we took a similar approach.
				The optimized energy levels used to evaluate the \SIrange{300}{400}{\angstrom}
				range were the upper and lower energy levels of most of the transitions in the
				\SIrange{900}{1200}{\angstrom} range. We used these levels to calculate Ritz wavelengths
				to compare to the wavelengths reported by \citeauthor{Raassen-Ni-V-Short}. The comparison,
				shown in Figure \ref{Ritz-Rass-Mid-Range}, demonstrated that the calibration error trend seen
				in the \SIrange{1200}{1300}{\angstrom} range (shown in Figure \ref{Good-Invar-Graph})
				continued down towards \SI{1100}{\angstrom}. We corrected the wavelengths in the
				\SIrange{1100}{1200}{\angstrom} range by shifting down the wavelengths reported by
				\citeauthor{Raassen-Ni-V-Short} by \SI{11}{\milli\angstrom} (the average of the differences
				between the Ritz wavelengths and the wavelengths reported by \citeauthor{Raassen-Ni-V-Short}).
				We increased the uncertainty of these wavelengths by the standard deviation of the set of
				differences (\SI{12}{\milli\angstrom}). This \SI{12}{\milli\angstrom} correction was added
				to the original measurement uncertainty of each wavelength as a sum in quadrature. The
				wavelengths in Table \ref{Ni-V-Table} that have been corrected in this way are reported
				as the observed wavelengths with a mark (R3) in the note column.


				The wavelengths in the \SIrange{900}{1100}{\angstrom} range and the
				\SIrange{1300}{1400}{\angstrom}, similar to those in the
				\SIrange{300}{400}{\angstrom} range, did not demonstrate any systematic errors, so
				we have given them in Table \ref{Ni-V-Table} as the original values given by
				\citeauthor{Raassen-Ni-V-Long} with a mark (R4) in the note column.

		\subsection{Intensity}\label{Intensity-Results}

			The line intensities in column five of Table \ref{Ni-V-Table} were taken from
			our spectra when available. If an accurate intensity could not be determined
			from our spectra due to issues with fitting the line profile, which could occur
			as a result of blending or having a weak line on the shoulder of stronger lines,
			then the line includes a characteristic mark in Table \ref{Ni-V-Table} indicating
			an unreliable intensity value.

			As not all of the lines reported by \citeauthor{Raassen-Ni-V-Long} and
			\citeauthor{Raassen-Ni-V-Short} were measured in this work, the line intensities
			reported in Table \ref{Ni-V-Table} are on two scales. Lines that have updated
			intensity measurements through this work are on the calibrated scale, while lines
			that were not measured in this work are reported on the original scale set by
			\citeauthor{Raassen-Ni-V-Long} and \citeauthor{Raassen-Ni-V-Short}. Table \ref{Ni-V-Table}
			includes a clear marker in the note column on each entry to indicate if the line
			intensity is from the original (noted as R) or updated scale (noted as W).

		\subsection{$\log(gf)$}\label{log(gf)}

\begin{figure}

				\centering

				\includegraphics[width=\columnwidth]{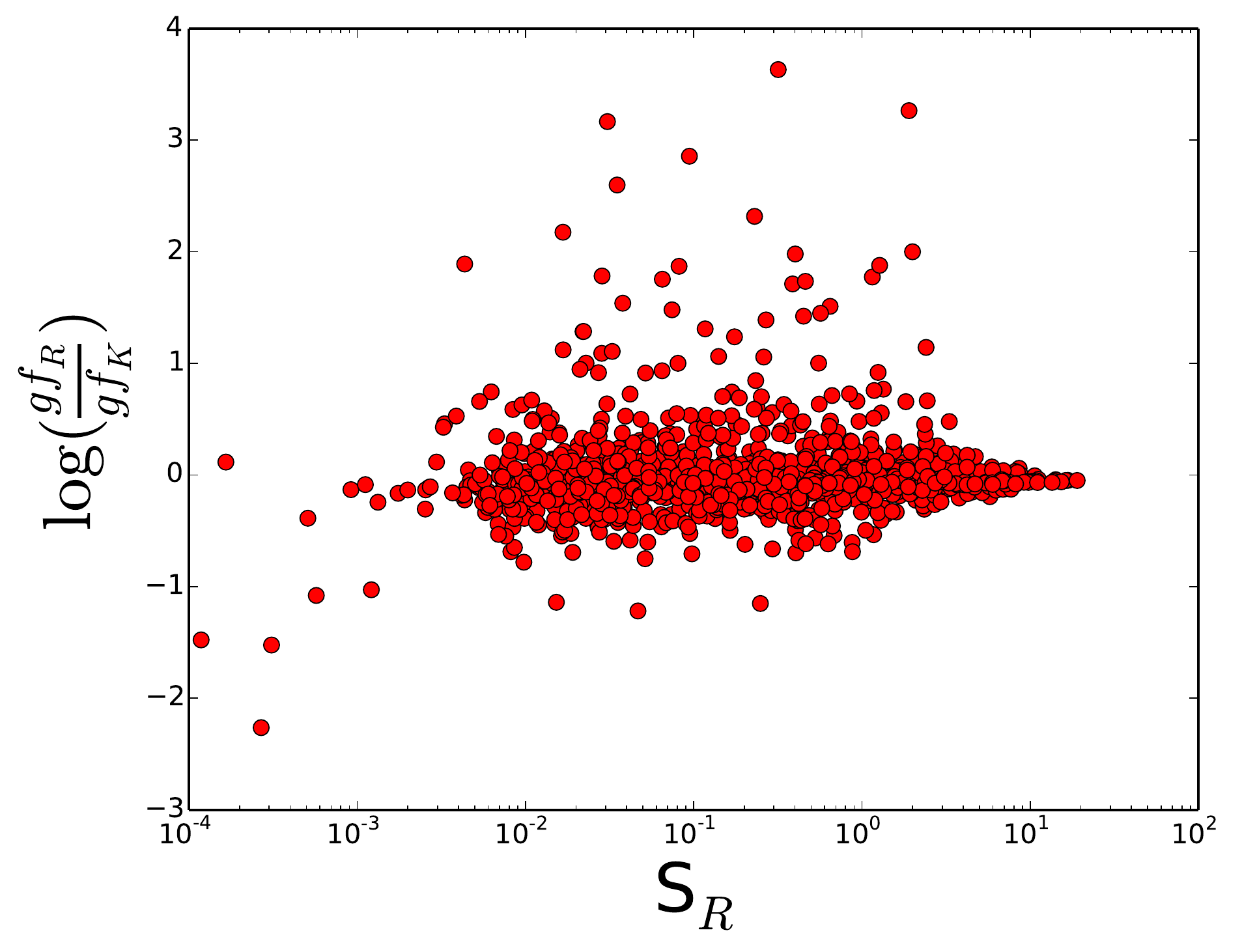}

				\caption
				{
					Difference between the $\log(gf)$ values reported by \citet{Raassen-Calc} ($\log(gf_{R})$)
					and \citet{Kurucz-Calc} ($\log(gf_{K})$) as a function of line strengths calculated
					using values from \citet{Raassen-Calc} ($S_{R}$). The line strengths are given in atomic
					units defined by eqautaion \ref{au-def}.\label{log(gf)-Graph}
				}

			\end{figure}

			The $\log(gf)$ values presented in Table \ref{Ni-V-Table} are the result
			of detailed calculations carried out by \citet{Raassen-Calc}. The accuracy
			of those $\log(gf)$ values was assessed by comparing them to the $\log(gf)$
			values calculated by \citet{Kurucz-Calc}. Figure \ref{log(gf)-Graph} presents
			the difference of the two sets as a function of the calculated line strength
			values given in atomic units (a.u.):
			\begin{equation}
			\label{au-def}
				a_{0}^{2}e^{2}=\SI{2.729e-48}{\meter\squared\coulomb\squared}
			\end{equation}
			where $a_{0}$ is the Bohr radius and $e$ is the electric charge.
			The plot in Figure \ref{log(gf)-Graph} has a standard deviation
			of $0.3$.


			Historically, the calculations provided by \citet{Raassen-Calc} have been far
			more accurate than other calculations \citep{Fe-II-Calc} and so the
			uncertainties for the $\log(gf)$ values provided by \citet{Raassen-Calc} can
			be roughly estimated by taking the standard deviation of the difference of
			the two sets of $\log(gf)$ values as a function of the line strengths calculated
			by \citet{Kurucz-Calc}. This results in a conservative upper limit for
			the $\log(gf)$ uncertainties reported in Table \ref{Ni-V-Table}. Ultimately,
			the uncertainties for the $\log(gf)$ values were broken down into three levels of
			quality based on the line strength. The weakest lines ($S$ $\leq$ 5 a.u.) have the
			lowest rating (uncertainty $> \SI{50}{\percent}$), moderate lines ( 5 a.u. $< S \leq$ 10 a.u.) have the
			middle rating (uncertainty $\leq \SI{18}{\percent}$), and the strongest lines ($S >$ 10 a.u.) have
			the highest rating (uncertainty $\leq \SI{7}{\percent}$). These different uncertainty levels are
			given in column seven of Table \ref{Ni-V-Table}.

		\subsection{Level Optimization}\label{LOPT}

			\begin{deluxetable*}{ccccccc}[htb!]
				\tabletypesize{\scriptsize}
				\tablecaption{Energy Levels of Ni V\label{Level-Opt-Table}}
				\tablehead
				{
					\colhead{Configuration} &
					\colhead{Term\,$^{a}$} &
					\colhead{J} &
					\colhead{Energy} &
					\colhead{$u_{E}$\,$^{b}$} &
					\colhead{Number of Transitions} \\
					\colhead{} &
					\colhead{} &
					\colhead{} &
					\colhead{{(}\si{\cm}$^{-1}${)}} &
					\colhead{{(}\si{\cm}$^{-1}${)}} &
					\colhead{Determining Level}
				}

				\startdata
					3d$^6$ & $^5$D     & 4 &      0.00 & 0.00 & 33 \\
					3d$^6$ & $^5$D     & 3 &    889.61 & 0.29 & 37 \\
					3d$^6$ & $^5$D     & 2 &   1489.82 & 0.32 & 32 \\
					3d$^6$ & $^5$D     & 1 &   1871.38 & 0.35 & 25 \\
					3d$^6$ & $^5$D     & 0 &   2057.52 & 0.46 & 10 \\
					3d$^6$ & $^3$P$_2$ & 2 & 26,152.49 & 0.38 & 24 \\
					3d$^6$ & $^3$H     & 6 & 27,111.40 & 0.33 & 28 \\
					3d$^6$ & $^3$H     & 5 & 27,578.61 & 0.32 & 40 \\
					3d$^6$ & $^3$H     & 4 & 27,858.94 & 0.32 & 41 \\
					3d$^6$ & $^3$P$_2$ & 1 & 28,697.33 & 0.40 & 20 \\
					3d$^6$ & $^3$F$_2$ & 4 & 29,123.90 & 0.30 & 49 \\
				\enddata

				\tablecomments{Table \ref{Level-Opt-Table} is published in its entirety in the electronic edition of the {\it Astrophysical Journal}.}

					\tablenotetext{a}{Note: * -- Odd Parity.}
					\tablenotetext{b}{Estimate of one standard uncertainty.}

			\end{deluxetable*}

				We have optimized the energy levels of Ni V with the set of critically evaluated
				wavelengths described in section \ref{Ni-V-Wave} as was done by \citet{Kramida-Ritz}
				for Fe V. The optimization process was done with the Level Optimization program (LOPT)
				created by \citet{Kramida-LOPT}. LOPT was also used to generate Ritz wavelengths.
				The Ritz wavelengths we have derived have uncertainties that are typically smaller
				than their experimentally measured counterparts.

				LOPT uses the inverse square of the wavelength uncertainty (column two of Table
				\ref{Ni-V-Table}) to weight each transition in the optimization and decreases
				the weight of all multiply classified lines. Since many lines in this
				optimization were multiply classified, the $gf$ values, taken from the $\log(gf)$
				values discussed in section \ref{log(gf)}, were used as additional weights for
				multiply classified lines. This was rarely used as almost all levels could be determined
				by lines that were not multiply classified. In the cases where levels did not depend on
				multiply classified lines, the wavelength uncertainty of those multiply classified
				lines was increased to \SI{20}{\milli\angstrom} in the LOPT input file so that the
				multiply classified lines would not impact the calculated energy levels, but would
				be included in the optimization files in order to determine their corresponding
				Ritz wavelength.

				The Ritz wavelengths, along with their estimated standard uncertainties,
				are reported in Table \ref{Ni-V-Table}. The optimized energy levels, their
				uncertainties, and their classifications are reported in Table
				\ref{Level-Opt-Table}. The level uncertainty given in column five of Table
				\ref{Level-Opt-Table} is one standard uncertainty with respect to the ground
				level. The number of transitions defining a level is included in Table
				\ref{Level-Opt-Table} in addition to the level uncertainty in order to give a
				full representation of each optimized level.


	\section{Conclusions}

		The original motivation behind this work was ultimately to improve
		the quality of astrophysical assessments of the fine structure
		constant. The work presented here supports the wavelength
		evaluation presented by both \citet{Ekberg-Fe-V} and
		\citet{Kramida-Ritz}. With the newly established laboratory and
		Ritz wavelengths for Ni V the results of \citet{Berengut-Alpha-Dwarf}
		can be revisited and improved upon. The Ni V systematic calibration error
		that is identified in this work can account for many of the inconsistencies
		between the iron and nickel data.

		The comprehensive compilation of data presented
		in this work has a wide range of applications from astronomy to fusion research.
		In connection to white dwarf stars, it can be used to further develop more accurate
		models of hot white dwarf atmospheres with non-LTE conditions and to determine
		relative abundances \citep{Werner-Dwarf-Abundances, Preval-Dwarf-Atmosphere}.


\acknowledgments

		J. W. Ward would like to thank the NIST Summer Undergraduate
		Research Fellowship Program for providing funding for his contributions
		to this research. J. W. Ward would also like to thank Trey Porto for his
		contributions to this project as a reader and university advisor. We would
		also like to thank Joseph Reader and Albert Henins for their guidance and
		assistance with the instrumentation for this work. This work was partially
		funded by the National Aeronautics and Space Administration of the USA,
		Grant NNH17AE08I.



\appendix

	\section{Supplementary Table}\label{Appendix-A}

		\begin{deluxetable}{ccccc}[h!]

			\tabletypesize{\scriptsize}
			\tablecaption{Comparison of Fe V DataFigures\label{Fe-V-Table}}
			\tablehead
			{
				\colhead{$\lambda_{obs}$\,\,{(}\si{\angstrom}{)}\,\,$^{a}$} &
				\colhead{$u_{obs}$\,\,{(}\si{\milli\angstrom}{)}\,\,$^{b}$} &
				\colhead{$\lambda_{E}$\,\,{(}\si{\angstrom}{)}\,\,$^{c}$} &
				\colhead{$\lambda_{k}$\,\,{(}\si{\angstrom}{)}\,\,$^{d}$} &
				\colhead{$u_{k}$\,\,{(}\si{\milli\angstrom}{)}\,\,$^{b}$}
				}

				\startdata
					1234.642 & 2.4 & 1234.648 & 1234.6455 & 2.2 \\
					1280.471 & 3.1 & 1280.471 & 1280.4678 & 2.1 \\
					1284.107 & 2.4 & 1284.109 & 1284.1080 & 1.7 \\
					1285.920 & 2.6 & 1285.918 & 1285.9150 & 2.1 \\
					1288.164 & 2.4 & 1288.169 & 1288.1681 & 1.8 \\
					1293.377 & 2.4 & 1293.377 & 1293.3826 & 1.8 \\
					1297.544 & 2.4 & 1297.547 & 1297.5453 & 1.8 \\
					1300.605 & 2.4 & 1300.608 & 1300.6095 & 1.7 \\
					1311.828 & 2.4 & 1311.828 & 1311.8290 & 3.0 \\
					1320.412 & 2.4 & 1320.410 & 1320.4116 & 2.0 \\
					\enddata

			\tablecomments{Table \ref{Fe-V-Table} is published in its entirety in the electronic edition of the {\it Astrophysical Journal}.}

				\tablenotetext{a}{Wavelengths measured in this report.}
				\tablenotetext{b}{One standard uncertainty of the wavelength value in the previous column.}
				\tablenotetext{c}{Wavelengths as reported by \citet{Ekberg-Fe-V}}
				\tablenotetext{d}{Wavelengths as reported by \citet{Kramida-Ritz}}

		\end{deluxetable}


\clearpage

\end{document}